# Ordinal Boltzmann Machines for Collaborative Filtering

**Tran The Truyen, Dinh Q. Phung, Svetha Venkatesh**
Department of Computing
Curtin University of Technology
Kent St, Bentley, WA 6102, Australia
*{t.tran2,d.phung,s.venkatesh}@curtin.edu.au*

## Abstract

Collaborative filtering is an effective recommendation technique wherein the preference of an individual can potentially be predicted based on preferences of other members. Early algorithms often relied on the strong locality in the preference data, that is, it is enough to predict preference of a user on a particular item based on a small subset of other users with similar tastes or of other items with similar properties. More recently, dimensionality reduction techniques have proved to be equally competitive, and these are based on the co-occurrence patterns rather than locality. This paper explores and extends a probabilistic model known as Boltzmann Machine for collaborative filtering tasks. It seamlessly integrates both the similarity and co-occurrence in a principled manner. In particular, we study parameterisation options to deal with the ordinal nature of the preferences, and propose a joint modelling of both the user-based and item-based processes. Experiments on moderate and large-scale movie recommendation show that our framework rivals existing well-known methods.

## 1 INTRODUCTION

Collaborative filtering is based on the idea that we can predict preference of an user on unseen items by using preferences already expressed by the user and others. For example, if we want to predict how much the user likes a particular movie we may look for similar users who have rated the movie before (Resnick *et al.*, 1994). Alternatively, the rating for this new movie can be based on ratings of other similar movies that the user has watched (Sarwar *et al.*, 2001). This *similarity-based* approach relies on the strong locality in the neighbourhood of highly correlated users or items. More recent development has suggested that *dimensionality reduction* techniques like SVD (Salakhutdinov *et al.*, 2007), PLSA (Hofmann, 2004) and LDA (Marlin, 2004) are also competitive. The idea is to assume a low dimensional representation of rating data, which, once learnt, can be used to generate unseen ratings. Unlike the similarity-based approach, this does not assume any locality in the data.

In this paper, we take the view that these approaches are complementary and they address different aspects of the user's preferences. Specifically, we explore the application of an undirected graphical model known as Boltzmann Machines (BMs) (Ackley *et al.*, 1985) for the problem. The strength of BMs comes from the capacity to integrate the *latent aspects* of user's preferences as well as the *correlation* between items and between users. The undirected nature of the model allows flexible encoding of data, and at the same time, it supports inference and learning in an principled manner. For example, the model supports missing ratings and joint predictions for a set of items and users. It provides some measure of *confidence* in each prediction made, making it easy to assess the nature of recommendation and rank results. With the hidden variables we can project user's preferences and item ratings onto a latent low dimensional space for further processing. Note that its probabilistic integration differs from the current practice of blending multiple independent models (Koren, 2008).

Importantly, we go beyond the standard BMs in a number of ways. Firstly, we explore various parameterisations to deal with the *ordinal* nature of ratings (e.g. if the true rating is 3 stars in a 5-star scale, then predicting 4 stars is preferred to predicting 5 stars). The standard discrete graphical models, on the other hand, count both the predictions as errors. One way to deal with this issue is to approximate them by continuous variables as done in (Hofmann, 2004) but this is only meaningful for numerical ratings. Secondly, previous BMs generally assume that each subset of observational variables are generated from an hidden process of the same type, and the data comes with a set of *i.i.d* instances. In collaborative filtering, on the other hand, it is much more plausible to assume that observed ratings are *co-generated* by both the user-based and item-based processes. As a result, the data instances are no longer independently and identically distributed. To deal with this, we propose to integrate data instances into a single BM, in which, every rating is associated with both the user-based and item-based processes. Further, this paper studies approximate learning strategies for large BMs, including Contrastive Divergence (Hinton, 2002) (CD), a *structural* extension to Pseudo-Likelihood (Besag, 1975) (PL), and the combination of CD and PL for the joint model.

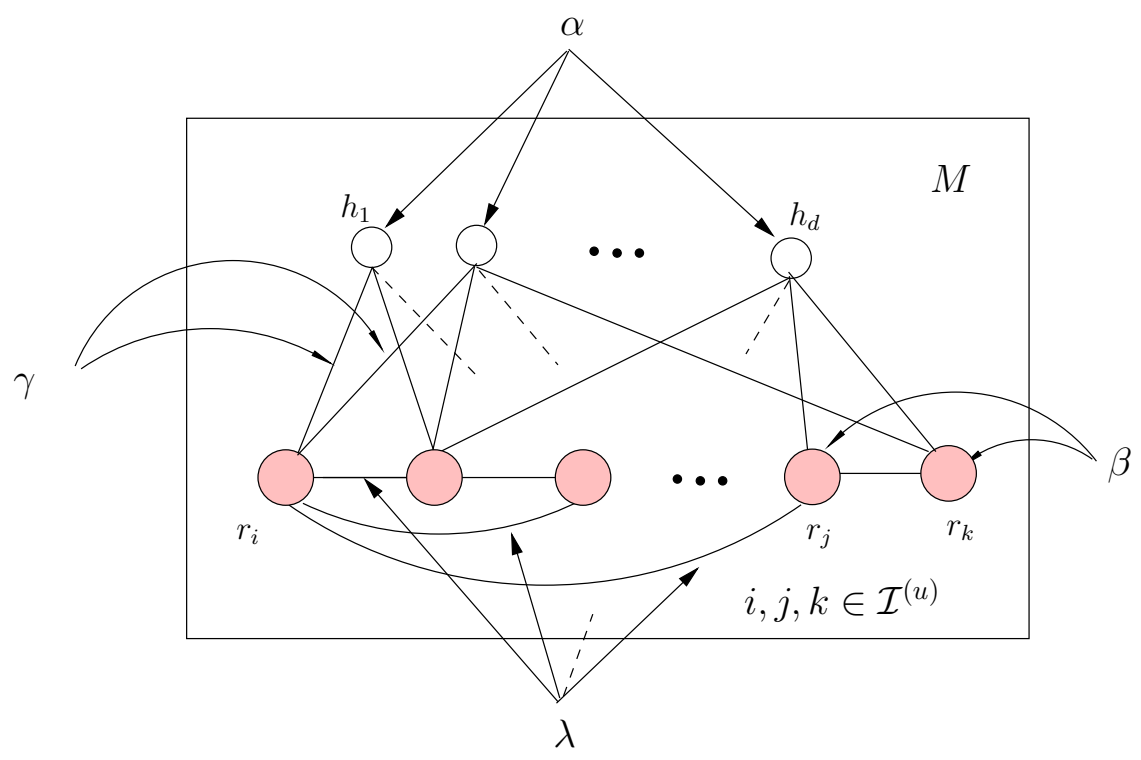

Figure 1: Plate graphical representation for user-centric modelling with Boltzmann machine. For a user $u$, the top layer represents the Boolean-valued hidden variables, the bottom layer represents the set of observed ratings for that user. The parameters are shared among all $M$ users.

## 2 USER-CENTRIC MODELLING

Denote by $\mathcal{U} = \{1, 2, \ldots, M\}$ the set of $M$ users, and $\mathcal{I} = \{1, 2, \ldots, K\}$ the set of $K$ items in the recommendation system of interest. Let us further denote by$\mathcal{S}$ the set of values a user can rate (e.g., $\mathcal{S} = \{1, 2, 3, 4, 5\}$ in the discrete case or $\mathcal{S} = [0, 1]$ in the continuous case). We use $u$ throughout this paper to index the user and $i$ to index the item. Let $\mathcal{I}^{(u)}$ be the set of indices of items rated by user $u$. Typically, the size of $\mathcal{I}^{(u)}$ is much smaller than the total number of items in the database (i.e. $|\mathcal{I}^{(u)}| \ll K$) because each user usually votes for only a small subset of items.

In this section, we first present probabilistic modelling from a single user perspective using Boltzmann machines (BMs). A user-centric BM in our view is an undirected graphical model representing user information and the set of associated rated items. A graphical model representation is shown in Fig. 1. There are two components in the model:

- a *hidden* layer to capture the latent aspects of a user modelled by a $d$-dim binary random vector variable $\boldsymbol{h} = (h_1, h_2, ..., h_d)$, and
- a *visible* layer representing ratings on different items observed for this user, captured by a random vector $\boldsymbol{r}^{(u)} = (r_i)_{i \in \mathcal{I}^{(u)}}$. Each element variable $r_i$ receives values in the set $\mathcal{S}$.

For the sake of understanding, we consider here discrete ratings where $\mathcal{S}$ is a finite discrete set and leave the case of continuous-valued ratings to the Appendix A.2. For clarity, we will drop explicit mention of user index $u$ and the membership relation $i \in \mathcal{I}^{(u)}$ and reinstate them whenever confusion may arise.

In the extreme view, the user-centric model should have represented both rated and non-rated items, treating all non-rated items as hidden variables at the bottom layer. However, since we do not have the knowledge of which items the user will rate in the future while the number of items is typically large (at the scale of millions in real-world scenarios), it will be impractical to include all the unknown ratings into the model. Our strategy is to limit to only known ratings at training time and gradually introduce an additional unknown rating at the prediction time as an unobserved variable subject to be inferred.

To parameterise our model, we first consider two additional kinds of features extracted from the set of ratings: for each rating $r_i$ we extract a vector of features $\{f_a(r_i)\}_{a=1}^{A}$, and for each rating pair$\{r_i, r_j\}$ a feature vector $\{f_b(r_i, r_j)\}_{b=1}^{B}$. While $f_a(r_i)$ captures some intrinsic property of the item $i$ and the rating $r_i$, $f_b(r_i, r_j)$ encodes correlation between the two item $i, j$ and their corresponding ratings. Four types of parameters are introduced (c.f. Fig. 1): each hidden unit $h_k$ is parameterised with $\alpha_k$, each feature $f_a(r_i)$ at the rating $r_i$ with $\beta_{ia}$, each pair $(h_k, f_a(r_i))$ with $\gamma_{kia}$, and each item-to-item correlation feature $f_b(r_i, r_j)$ with $\lambda_{ijb}$. For the user $u$, the model state negative energy is now ready to be defined as

$$-E^{(u)}(\boldsymbol{h}, \boldsymbol{r}) = \begin{cases} \sum_{1 \le k \le d} \alpha_k h_k + \sum_{i \in \mathcal{I}^{(u)}, a} \beta_{ia} f_a(r_i) \\ + \sum_{i \in \mathcal{I}^{(u)}, k, a} \gamma_{ika} h_k f_a(r_i) \\ + \sum_{i, j \in \mathcal{I}^{(u)}; i \neq j} \sum_b \lambda_{ijb} f_b(r_i, r_j) \end{cases}$$

where $\alpha_k, \beta_{ia}, \gamma_{ika}$ and $\lambda_{ijb}$ are model parameters which are *shared* among users as shown to be outside the plate in Fig.1 . Finally, the user-centric model distribution follows

$$P^{(u)}(\boldsymbol{h}, \boldsymbol{r}) = \frac{1}{Z(u)} \exp\{-E^{(u)}(\boldsymbol{h}, \boldsymbol{r})\} \qquad (1)$$

where $Z(u) = \sum_{\boldsymbol{h},\boldsymbol{r}} \exp\{-E^{(u)}(\boldsymbol{h},\boldsymbol{r})\}$ is the normalising constant. Denote by $h_{k,1}$ the assignment $h_k = 1$. The conditional distributions are (again we drop user index $u$)

$$P(h_{k,1} \mid \boldsymbol{r}) = \left[1 + \exp\left\{-\alpha_k - \sum_{ia} \gamma_{ika} f_a(r_i)\right\}\right]^{-1}$$
$$P(r_i | \boldsymbol{r}_{\neg i}, \boldsymbol{h}) \propto \exp\{I(r_i, \boldsymbol{h}) + J(r_i, \boldsymbol{r}_{\neg i})\}$$

where $\boldsymbol{r}_{\neg i}$ denote the set of ratings by the same user $u$ other than $r_i$, and

$$I(r_i, \boldsymbol{h}) = \sum_a \beta_{ia} f_a(r_i) + \sum_{ka} \gamma_{ika} f_a(r_i) h_k$$
$$J(r_i, \boldsymbol{r}_{\neg i}) = \sum_{j \neq i} \sum_b \lambda_{ijb} f_b(r_i, r_j)$$

For the purpose of dimensionality reduction we can use the vector $\{P(h_{k,1}|\boldsymbol{r})\}_{k=1}^d$ as a continuous representation of the user's preference.

### 2.1 ORDINAL FEATURES

In this paper we consider the case where user preferences are expressed in term of *ordinal* ratings, i.e., the set of rating values $\mathcal{S}$ is a set of $n$ ordinal values, and let us denote it by $\mathcal{S} = \{R_1, R_2, ...R_n\}$. A straightforward approach is to simply *ignore* the ordinal property and treat the ratings as *categorical* variables. In particular, the input bias feature can simply be an identity function $f_s(r_i) = \mathbb{I}[r_i \equiv R_s]$, and the correlation feature can be treated as the similarity between the two neighbour ratings $f_b(r_i, r_j) = \mathbb{I}[r_i \equiv r_j]$. Another way is to treat them as numerical values, for example, as random Gaussian variables (after appropriate preprocessing, see Appendix A.2 for a detailed treatment). However the shortcoming is that this treatment is only meaningful when such a numerical interpretation exists.

A better way is to exploit the *ordering* property: if the *true* rating by the user $u$ on item $i$ is $r_i = R_s$, then we would want to predict the rating as close to $R_s$ as possible. Denote by $R_{s'} \succ R_{s'+1}$ the preference of $R_{s'}$ to $R_{s'+1}$, the ordering of preferences when the true rating is $R_s$ can be expressed as

$$R_s \succ R_{s-1}.... \succ R_1$$
$$R_s \succ R_{s+1}... \succ R_n$$

It is essential to design *features* $\{f_a(r_i)\}_{a=1}^A$ to capture the information induced from these expressions. As $R_s$ split the set $\mathcal{S}$ into two subsets, we create one set of features corresponding to $s' < s$ and another set corresponding to $s" > s$, i.e. $f_{s'}^{down}(R_s) = (s' - s)\mathbb{I}[s' < s]$ and $f_{s"}^{up}(R_s) = (s" - s)\mathbb{I}[s" > s]$, respectively, where $\mathbb{I}[.]$ is the indicator function. For correlation between two items $(i, j)$, we can measure the distance between two corresponding ratings $r_i = R_s$ and $r_j = R_{s'}$ by user $u$, i.e. $f_b(r_i, r_j) = |s' - s|$.

### 2.2 LEARNING

Training data consists of rating values for input variables. Let us denote these *evidences* per user $u$ as $\bar{\boldsymbol{r}}^{(u)}$, to distinguish from the unspecified $\boldsymbol{r}^{(u)}$. Standard maximum likelihood learning maximises $\mathcal{L} = \sum_{u \in \mathcal{U}} \mathcal{L}(\bar{\boldsymbol{r}}^{(u)})$, where $\mathcal{L}(\bar{\boldsymbol{r}}^{(u)}) = \log P^{(u)}(\bar{\boldsymbol{r}}^{(u)})$. Let us drop the index $u$ for clarity and take the gradient with respect to model parameters yielding

$$\frac{\partial \mathcal{L}(\bar{\boldsymbol{r}})}{\partial \alpha_k} = P(h_{k,1}|\bar{\boldsymbol{r}}) - P(h_{k,1})$$
$$\frac{\partial \mathcal{L}(\bar{\boldsymbol{r}})}{\partial \beta_{ia}} = f_a(\bar{r}_i) - \sum_{r_i} P(r_i) f_a(r_i)$$
$$\frac{\partial \mathcal{L}(\bar{\boldsymbol{r}})}{\partial \gamma_{ika}} = P(h_{k,1}|\bar{\boldsymbol{r}}) f_a(\bar{r}_i) - \sum_{r_i} P(r_i, h_{k,1}) f_a(r_i)$$
$$\frac{\partial \mathcal{L}(\bar{\boldsymbol{r}})}{\partial \lambda_{ijb}} = f_b(\bar{r}_i, \bar{r}_j) - \sum_{r_i, r_j} P(r_i, r_j) f_b(r_i, r_j)$$

Generally, these gradients cannot be evaluated exactly. One method is to use Gibbs sampling to approximate the gradients. However, unbiased Gibbs sampling may take too much time to converge. We follow a sampling strategy called Contrastive-Divergence (CD) (Hinton, 2002), in that we start the sampling from the data distribution, and stop the random walks after a few steps. This certainly introduces bias, but it is enough to relax the distribution toward the true distribution, and more importantly, it is very efficient.

Another method is to utilise Pseudo-likelihood (PL) (Besag, 1975), and we approximate the model log-likelihood by

$$\mathcal{L}^{PL}(\bar{\boldsymbol{r}}) = \sum_{i \in \mathcal{I}^{(u)}} \log P(\bar{r}_i | \bar{\boldsymbol{r}}_{\neg i})$$

Note that in the original PL, there are no hidden variables, thus computing the local conditional distribution $P(\bar{r}_i|\bar{\boldsymbol{r}}_{\neg i})$ is easy. In our case, the pseudo-likelihood and its gradient can also be computed exactly and efficiently but the derivations are rather involved, and we leave the details in the Appendix A.1.

### 2.3 RATING PREDICTION

Once trained, the BMs can be used for predicting the preference of a user. Recall that unseen items are not modelled during training but will be added as an additional, unobserved node in the visible layer during testing[1]. The prediction on new item $j \notin \mathcal{I}^{(u)}$ is based on the MAP assignment[2]

$$r_j^* \quad = \quad \arg\max_{r_j} P(r_j|\bar{\boldsymbol{r}})$$

where $P(r_j^*|\bar{\boldsymbol{r}})$ is the measure of *prediction confidence*. Given $\bar{\boldsymbol{r}}$, the model structure is reduced to a tree with the root $r_j$ and leaves $\{h_k\}_{k=1}^d$. Thus $r_j^*$ can be evaluated in linear time. However, the computation is still expensive for online deployment. Here we propose to use a cheaper method, which is based on *mean-field* approximation:

$$P(r_j, \boldsymbol{h}|\bar{\boldsymbol{r}}) \approx Q(r_j|\bar{\boldsymbol{r}}) \prod_k Q(h_k|\bar{\boldsymbol{r}})$$

For simplicity, we fix $Q(h_k|\bar{\boldsymbol{r}}) = P(h_k|\bar{\boldsymbol{r}})$ based on the idea that previous information is rich enough to shape the distribution $Q(h_k|\bar{\boldsymbol{r}})$. Minimizing the Kullback-Leibler divergence between $P(r_j, \boldsymbol{h}|\bar{\boldsymbol{r}})$ and its approximation, we obtain $Q(r_j|\bar{\boldsymbol{r}}) \propto \exp(-E_Q(r_j, \bar{\boldsymbol{r}}))$, where

$$E_Q(r_j, \bar{\boldsymbol{r}}) = \begin{cases} -\sum_a \beta_{ja} f_a(r_j) \\ -\sum_k P(h_{k,1}|\bar{\boldsymbol{r}}) \sum_a \gamma_{jka} f_a(r_j) \\ -\sum_{i \in \mathcal{I}^{(u)}, b} \lambda_{ijb} f_b(\bar{r}_i, r_j) \end{cases} \quad (2)$$

This is equivalent to replacing the hard hidden assignment $h_k \in \{0, 1\}$ by a soft values $P(h_{k,1}|\bar{\boldsymbol{r}}) \in [0, 1]$. Finally, using $Q(r_j|\bar{\boldsymbol{r}})$ in place of $P(r_j|\bar{\boldsymbol{r}})$, the prediction is simplified as $r_j^* = \arg\min_{r_j} E_Q(r_j, \bar{\boldsymbol{r}})$.

Note that this mean-field approximation has the same linear complexity as the standard MAP, but it is numerically faster because the mathematical expression is more computationally primitive.

### 2.4 ITEM RANKING

In a recommendation system we are often interested in composing a recommendation list of items for each user. This is essentially a ranking problem, in that for a given set of candidate items, we need to provide a numerical score for each item and choose a top ranked items. In our BMs framework, adding a new item $j$ to the model will approximately reduce the model state energy by an amount of $E_Q(r_j, \bar{\boldsymbol{r}})$ (defined in Equation 2). Recall that the user likelihood in Equation 1 improves if the model state energy decreases, thus motivating us to use the $E_Q(r_j, \bar{\boldsymbol{r}})$ as the ranking score. Since we do not know exactly the state $r_j$, we resort to the (approximate) *expected energy decrease* instead $\Delta E_j = \sum_{r_j} Q(r_j|\bar{\boldsymbol{r}}) E_Q(r_j, \bar{\boldsymbol{r}})$.

[1] It may appear that adding new item can make the user model unspecified, but in fact, the item is already in the models of other users and its related parameters have been learnt.

[2] Alternatively, we can take the expected rating as the prediction $r_j^* = \sum_{r_j} P(r_j|\bar{\boldsymbol{r}}) r_j$.

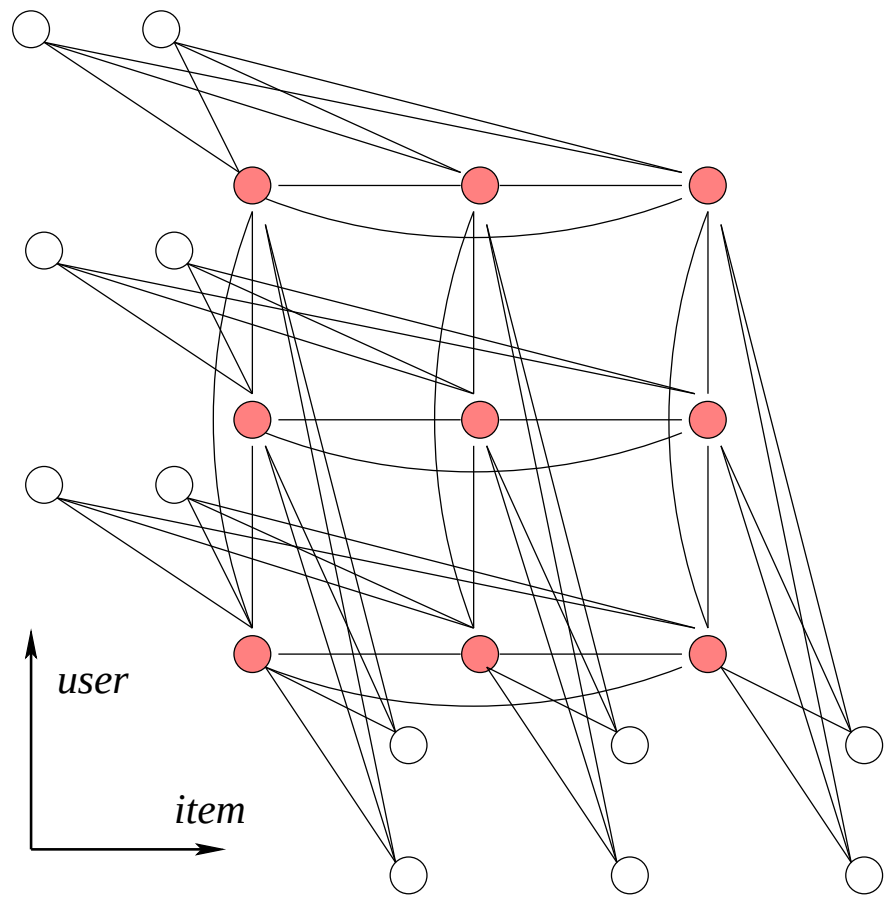

Figure 2: Joint modelling of users and items. Each row of filled nodes represents ratings per user; each row of empty nodes represents hidden tastes of an user; and each column of empty nodes depicts hidden features of an item.

## 3 JOINT MODELLING OF USERS AND ITEMS

In the previous section, we have assumed that ratings are generated by some user-centric process. Since users and items play an equal role in the data, we can alternatively assume that there exists some item-centric process that generates ratings. Thus we can alternatively model the ratings observed for each item instead of user in a manner similar to Section 2. The more plausible assumption, however, is that a rating is *co-generated* by both the user-centric and item-centric processes. This can be realised by combining these two modelling approaches into a single unified BM, as depicted in Figure 2.

More specifically, every user and item is modelled with its own hidden layer. Let $d'$ be the dimensionality of the hidden variables associated with items, there are $Md + Kd'$ hidden nodes in the joint model (every rating is associated with two hidden layers, one per the user and one per the item). The number of input nodes is the number of ratings in the whole database. Each input node corresponding to user $u$ and item $i$ is possibly connected to $|\mathcal{I}^{(u)}| + |\mathcal{U}^{(i)}| - 2$ other input nodes, where $\mathcal{U}^{(i)}$ denotes the set of all users who rate item $i$. Thus, the resulting BM is a probabilistic database that supports various inference tasks.

Denote by $\mathbf{r}$ and $\mathbf{h}$ respectively the set of all input variables (i.e., observed ratings) and all hidden variables of the entire model. The energy of the entire system is

$$E(\mathbf{r},\mathbf{h}) = \sum_{u\in\mathcal{U}} E^{(u)}(\boldsymbol{r}^{(u)},\boldsymbol{h}^{(u)}) + \sum_{i\in\mathcal{I}} E^{(i)}(\boldsymbol{r}^{(i)},\boldsymbol{h}^{(i)})$$

where

$$-E^{(i)}(\boldsymbol{r}^{(i)},\boldsymbol{h}^{(i)}) = \begin{cases} \sum_k \theta_k h_k + \sum_{u\in\mathcal{U}^{(i)},a} \eta_{ua} f_a(r_u) \\ + \sum_{u\in\mathcal{U}^{(i)},k,a} \nu_{uka} h_k f_a(r_u) \\ + \sum_{u,v\in\mathcal{U}^{(i)};u\neq v} \sum_b \omega_{uvb} f_b(r_u,r_v) \end{cases}$$

where $\theta,\eta,\nu,\omega$ are respective item-centric model parameters that play similar roles to $\alpha,\beta,\gamma,\lambda$ in user-centric models.

Let $\bar{\mathbf{r}}$ denote all assigned rating values in the training data. Since the model structure is complex, we look for decomposition to simplify computation. As ratings can be decomposed by either user indices or item indices, we appeal to *structured pseudo-likelihood* learning where we try to maximise the log pseudo-likelihood instead[3]:

$$\mathcal{L}^{PL}(\bar{\mathbf{r}}) = \frac{1}{2}\Big(\sum_{u\in\mathcal{U}} \log P(\bar{\boldsymbol{r}}^{(u)}|\bar{\mathbf{r}}_{\neg u}) + \sum_{i\in\mathcal{I}} \log P(\bar{\boldsymbol{r}}^{(i)}|\bar{\mathbf{r}}_{\neg i})\Big)$$

This objective function has a nice property that parameters associated with users and items are separated in corresponding components. Naturally, it suggests an *alternating* parameter updating strategy. Let us consider $P(\bar{\boldsymbol{r}}^{(u)} \mid \bar{\mathbf{r}}_{\neg u})$. Using the Markov property, $\bar{\mathbf{r}}_{\neg u}$ reduces to ratings by all neighbours of user $u$. Since each item rated by user $u$ has its own hidden variables and integrating out these variables in standard likelihood learning may be expensive (although feasible), we further propose a mean-field approximation approach, similar to that described in Section 2.3. More specifically, when we update parameter associated with user $u$, we considered the hidden layer for item $i$ observed with value $\{P(h^{(i)}_{k',1}|\bar{\boldsymbol{r}}^{(i)})\}^{d'}_{k'=1}$. The learning now reduces to that described in Section 2.2. The overall algorithm can be summarised as follows

- Loop until stopping criteria met:
  - Use $\{P(h^{(i)}_{k',1}|\bar{\boldsymbol{r}}^{(i)})\}^{d'}_{k'=1}$ as hidden values per item $i$, for all $i\in\mathcal{I}$. Fix item models parameters, update user model parameters by maximising $\sum_{u\in\mathcal{U}} \log P(\bar{\boldsymbol{r}}^{(u)}|\bar{\mathbf{r}}_{\neg u})$.
  - Use $\{P(h^{(u)}_{k,1}|\bar{\boldsymbol{r}}^{(u)})\}^{d}_{k=1}$ as hidden values per user $u$, for all $u\in\mathcal{U}$. Fix user model parameters, update item model parameters by maximising $\sum_{i\in\mathcal{I}} \log P(\bar{\boldsymbol{r}}^{(i)}|\bar{\mathbf{r}}_{\neg i})$.

[3]Note that there is a single distribution $P(\mathbf{r},\mathbf{h})$ for the whole data.

In the testing phase, we introduce a single node $r_{uj}$ to the model and compute $r^*_{uj} = \arg\max_{r_{uj}} P(r_{uj}|\bar{\mathbf{r}})$, which can be simplified further by noting that $P(r_{uj}|\bar{\mathbf{r}}) = P(r_{uj}|\bar{\boldsymbol{r}}^{(u)},\bar{\boldsymbol{r}}^{(j)})$. We can also make use of the mean-field approximation similar to that in Section 2.3. More specifically, we make use of all the conditional distributions of hidden variables $\{P(h^{(i)}_{k',1}|\bar{\boldsymbol{r}}^{(i)})\}^{d'}_{k'=1}$ for each item $i$ and $\{P(h^{(u)}_{k,1}|\bar{\boldsymbol{r}}^{(u)})\}^{d}_{k=1}$ for each user $u$, then compute the energy decrease as

$$E_Q(r_{uj},\bar{\mathbf{r}}) = \begin{cases} -\sum_a \beta_{ja} f_a(r_{uj}) - \sum_a \eta_{ua} f_a(r_u) \\ -\sum_k P(h^{(u)}_{k,1}|\bar{\boldsymbol{r}}^{(u)}) \sum_a \gamma_{jka} f_a(r_{uj}) \\ -\sum_{i\in\mathcal{I}^{(u)},b} \lambda_{ijb} f_b(\bar{r}_i, r_j) \\ -\sum_k P(h^{(j)}_{k,1}|\bar{\boldsymbol{r}}^{(j)}) \sum_a \nu_{uka} f_a(r_u) \\ -\sum_{v\in\mathcal{U}^{(j)}} \sum_b \omega_{uvb} f_b(\bar{r}_u, r_v) \end{cases}$$

# 4 EVALUATION

## 4.1 SETTING

We evaluate the proposed Ordinal BMs on two movie rating datasets. The first moderate dataset comes from the MovieLens project[4], consisting of 6040 users, 3043 items and approximately 1 million ratings. The second larger dataset is extracted from the Netflix challenge[5] in that the first 3000 items are used, resulting in 208,332 users, and 13.6 million ratings[6]. Ratings are integers in the 5-star scale. The two datasets include only those users who have rated more than 20 movies, and those movies rated by more than 20 users. For each user, roughly 80% of ratings is used for training and the rest is for evaluation.

We implement three variants of the BMs: the *categorical*, the *ordinal* and the *Gaussian*. For the Gaussian BMs, we need to normalise the ratings to obtain random numbers following the standard normal distribution $\mathcal{N}(0;1)$. To determine the connectivity at the input layers, we first compute the Pearson correlation between user pairs and item pairs as in standard similarity-based methods (e.g. see (Sarwar *et al.*, 2001)), and keep only positively correlated pairs. Then, for each user/item we choose the top 100 similar users/items to be his/its neighbourhood, ranked by the Peason correlation. The BMs results reported here are based on *one-step* CD learning as it is empirically faster than the pseudo-likelihood method without much difference in performance. Models are trained in an online fashion with block size of 100, learning rate of 0.1. Parameters associated with hidden variables are initialised by a random Gaussian $\mathcal{N}(0; 0.01)$.

[4]http://www.grouplens.org

[5]http://netflixprize.com

[6]This subset, although smaller than the original 100 millions set, is still larger than the largest non-commercial dataset current available from the MovieLens project.

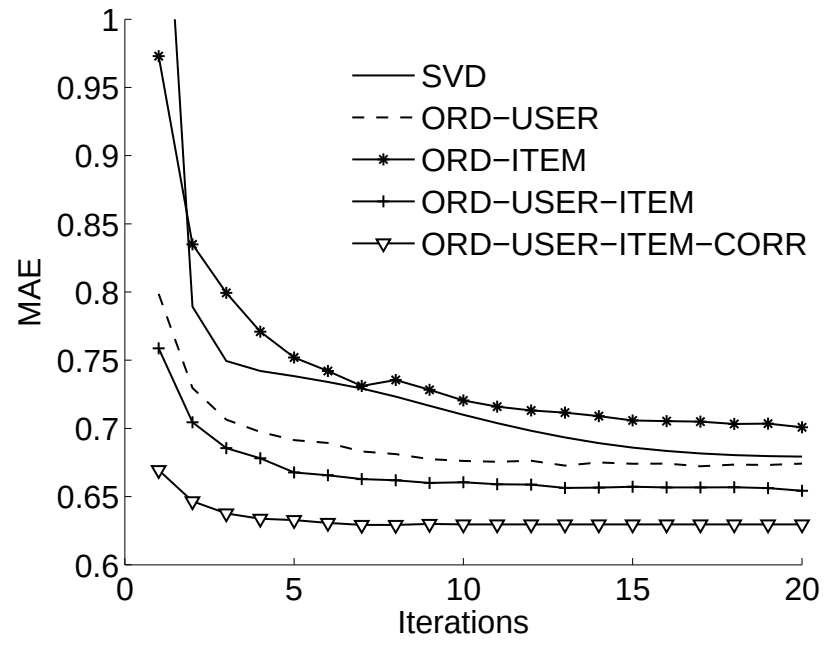

Figure 3: Rate prediction performance of Ordinal BM variants (MovieLens dataset). All model hidden sizes are fixed at 20. ORD-USER stands for an Ordinal BM per user, ORD-USER-ITEM for joint modelling *without* correlation at the input layer. ORD-USER-ITEM-CORR adds the correlation to ORD-USER-ITEM.

### 4.2 RATING PREDICTION

In the first set of experiments, we measure the performance of BM models on the *rating prediction* task (Section 2.3). For comparison, we implement the Singular Value Decomposition (SVD) for *incomplete* data (see, for example, (Salakhutdinov *et al.*, 2007) for a description). The SVD is currently one of the best methods for movie prediction[7]. The evaluation criterion is based on the popular *Mean Absolute Error* (MAE) measure, i.e. $\text{MAE}=\sum_{j=1}^{J}|r_j^* - \bar{r}_j|/J$.

Figure 3 shows the learning curves of BMs variants in comparison with the SVD, all evaluated on the 1M MovieLens dataset. The size of BM hidden layers and the rank of SVD are fixed at 20. The figure clearly demonstrates the positive effect of joint modelling, as well as of the integration of the dimensionality reduction and correlation. More importantly, the resultant model outperforms the SVD.

To investigate the role of hidden variables, we fix the number of iterations to 20 and run BMs variants under different hidden sizes *without* the correlation in the input layer, and the results are reported in Figure 4. Generally the performance (except for the categorical BMs) increases as more hidden units are introduced, but at a slow rate after 30 units.

The Netflix subset is characteristically different from the MovieLens dataset, in that there are far more users than items, thus it is not practical to include correlations between users (the number of correlation parameters is $M^2$, where $M$ is number of users). The results are reported in Table 1, which once again demonstrate the advantage of the proposed Ordinal BMs.

[7]It appears that all leaders in the Netflix competition use some forms of SVD.

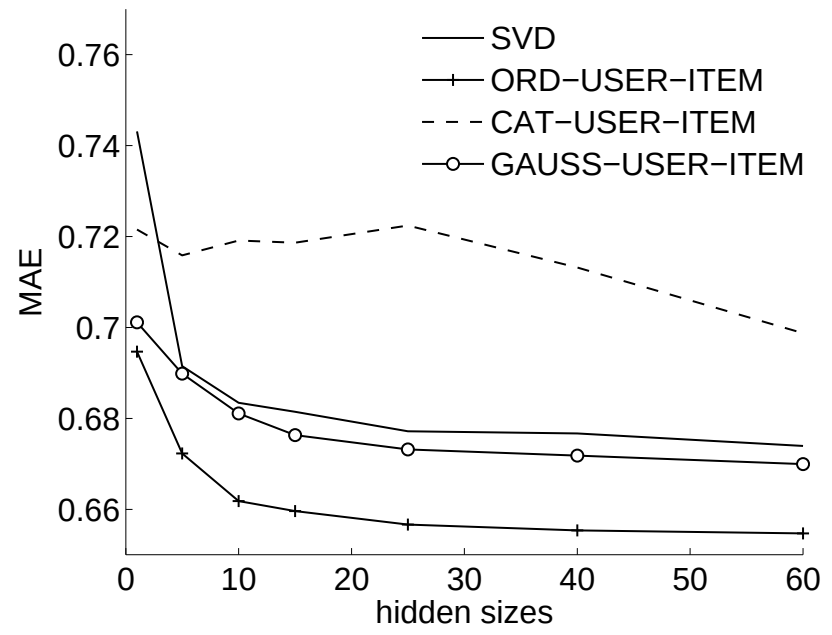

Figure 4: Rate prediction performance of Ordinal BM variants (MovieLens dataset) as a function of hidden dimensionality *without* correlation at the input layer. CAT-USER-ITEM stands for *categorical* joint modelling and GAUSS-USER-ITEM for *Gaussian*.

| | $K$=5 | $K$=20 | $K$=50 |
|---|---|---|---|
| SVD | 0.690 | 0.684 | 0.685 |
| CAT-USER | 0.766 | 0.693 | 0.682 |
| GAUSS-USER | 0.722 | 0.694 | 0.694 |
| ORD-USER | 0.707 | 0.663 | 0.649 |
| ORD-USER-ITEM | 0.678 | 0.649 | 0.645 |
| CAT-USER-CORR | 0.697 | 0.675 | 0.669 |
| GAUSS-USER-CORR | 0.687 | 0.687 | 0.689 |
| ORD-USER-CORR | 0.660 | 0.636 | 0.642 |
| ORD-USER-CORR-ITEM | **0.648** | **0.635** | **0.635** |

Table 1: Rating prediction on Netflix subset, measured in MAE. CAT-USER-CORR stands for Categorical BMs for individual users with input correlations, and ORD-USER-CORR-ITEM for joint modelling of users and items but considering only correlations in the user-based models.

### 4.3 ITEM RANKING

In the second set of experiments, we evaluate the Ordinal BMs for the *item ranking* task (Section 2.4). Recall that we first need a set of candidate items for each user. Here we use the Pearson similarity between users, that is, for each user $u$, we select 50 most similar users and then collect the items those users have previously rated. These items, except for those previously rated by user $u$, are the candidates. For comparison, we evaluate the Ordinal BMs against a baseline *popularity* method, in which importance of a candidate is based on the number of times it is rated by the neighbour users. Methods are tested on the MovieLens dataset only since the Netflix data is not suitable for computing user-based correlations. The evaluation criteria includes the standard recall/precision measures, and the *ranking utility* adapted from (Breese *et al.*, 1998). The utility is based on the assumption that the value of a recommendation, if it interests a user, is reduced exponentially with its position down the list. More specifically, for all recommended items that appear in the test set for user $u$, the ranking utility is computed as $\pi_u = \sum_p 2^{-(p-1)/(\alpha-1)}$, where $p$ is the position of the item in the recommendation list, and $\alpha > 0$ is the interest '*half-life*'. The overall ranking utility is then computed as

$$\pi = 100 \frac{\sum_u \pi_u}{\sum_u \pi_u^{\max}}$$

where $\pi_u^{\max} = \sum_{p'=1}^{T_u} 2^{-(p'-1)/(\alpha-1)}$ with $T_u$ be the size of the test set for user $u$. As suggested in (Breese *et al.*, 1998), we choose $\alpha = 5$. Figure 5 depicts the performance of the joint Ordinal BM, which clearly shows its competitiveness against the baseline. Hidden variables seem to play little role in this kind of inference, hence we report only result of model with input correlations and leave the issue for future investigation.

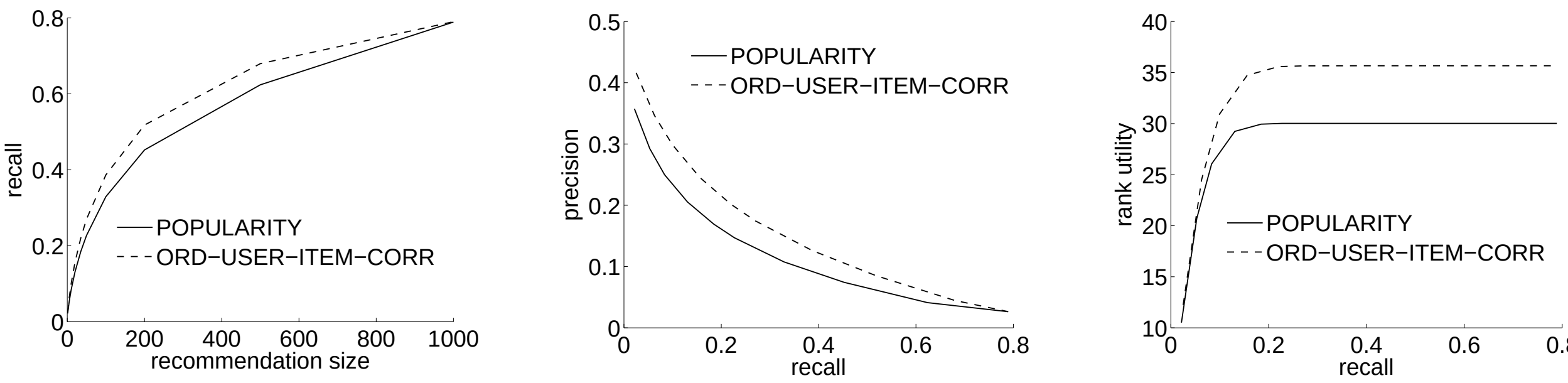

Figure 5: Ranking performance of the Ordinal BMs (ORD-USER-ITEM-CORR) against the baseline popularity-based (POPULARITY), on the 1M MovieLens data.

# 5 RELATED WORK

The Boltzmann Machines explored in this paper are more general that the original proposal in (Ackley *et al.*, 1985) due to the use of general exponential family instead of binary variables, in the same way that the Harmoniums (Welling *et al.*, 2005) generalises the Restricted BMs (e.g. see (Salakhutdinov *et al.*, 2007)). The work in (Salakhutdinov *et al.*, 2007) applies Restricted BMs for collaborative filtering but it is limited to individual modelling of users and categorical variables.

Other graphical models have been employed for collaborative filtering in a number of places, including Bayesian networks (Breese *et al.*, 1998) and dependency networks (Heckerman *et al.*, 2001). The BMs differ from Bayesian networks in that BMs are undirected models which Bayesian networks are directed. Our method resembles dependency networks when *pseudo-likelihood* (Besag, 1975) learning is employed and no hidden variables are modelled, but dependency networks are generally inconsistent.

The dimensionality reduction capacity of the BMs is shared by other probabilistic models, including mixture models, probabilistic latent semantic analysis (PLSA) (Hofmann, 2004) and latent Dirichlet allocation (LDA) (Marlin, 2004). These are all directed graphical models while the BMs are undirected. Machine learning (Billsus and Pazzani, 1998; Basu *et al.*, 1998; Basilico and Hofmann, 2004) has also been successfully applied to the collaborative filtering problem. The method maps the recommendation into a classification problem that existing classifiers can solve. The map typically considers each user or each item as an independent problem, and ratings are training instances.

# 6 CONCLUSION

We have presented Boltzmann machines for collaborative filtering tasks. BMs are an expressive framework to incorporate various aspects of the data, including the low dimensional representation of item/user profiles and the correlation between items/users. We study parameterisations for handling the ordinal nature of ratings, and propose the integration of multiple BMs for joint modelling of user-based and item-based processes. We empirically shown that BMs are competitive in the movie recommendation problem.

This work can be furthered in a number of ways. First we need to handle incremental parameter updating when new users or items are available. The second issue is learning the structure of the BMs, including determining the number of hidden units, and and connectivity in the input layer. And third, the model should be extended to incorporate external information like user profiles and item contents.

# A APPENDIX

## A.1 PSEUDO-LIKELIHOOD FOR THE DISCRETE BMs

Denote by

$$\begin{aligned}\phi_k(h_k) &= \exp\{\alpha_k h_k\}\\ \phi_i(r_i) &= \exp\{\sum_a \beta_{ia} f_a(r_i)\}\\ \psi_{ik}(h_k, r_i) &= \exp\{\sum_a \gamma_{ika} h_k f_a(r_i)\}\\ \psi_{ij}(r_i, r_j) &= \exp\{\sum_b \lambda_{ijb} f_b(r_i, r_j)\}\end{aligned}$$

Let us define the joint potential of the system $\Phi(\boldsymbol{h}, \boldsymbol{r})$ as

$$\left[\prod_k \phi_k(h_k)\right]\left[\prod_i \phi_i(r_i)\right]\left[\prod_{i,k} \psi_{ik}(h_k, r_i)\right]\left[\prod_{i,j} \psi_{ij}(r_i, r_j)\right]$$

In pseudo-likelihood (PL) learning we need to optimise the following objective function

$$\mathcal{L}^{PL}(\bar{\boldsymbol{r}}) = \sum_i \log P(\bar{r}_i|\bar{\boldsymbol{r}}_{\neg i})$$

where

$$P(\bar{r}_i|\bar{\boldsymbol{r}}_{\neg i}) = \frac{\sum_{\boldsymbol{h}} \Phi(\bar{r}_i, \bar{\boldsymbol{r}}_{\neg i}, \boldsymbol{h})}{\sum_{r_i}\sum_{\boldsymbol{h}} \Phi(r_i, \bar{\boldsymbol{r}}_{\neg i}, \boldsymbol{h})} = \frac{Z(\bar{r}_i|\bar{\boldsymbol{r}}_{\neg i})}{Z(\bar{\boldsymbol{r}}_{\neg i})}$$

where $Z(r_i|\bar{\boldsymbol{r}}_{\neg i}) = \sum_{\boldsymbol{h}} \Phi(r_i, \bar{\boldsymbol{r}}_{\neg i}, \boldsymbol{h})$ and $Z(\bar{\boldsymbol{r}}_{\neg i}) = \sum_{r_i} Z(r_i|\bar{\boldsymbol{r}}_{\neg i})$. Expanding $Z(r_i|\bar{\boldsymbol{r}}_{\neg i})$ and note that all potentials associated with $h_k = 0$ become 1, we obtain

$$Z(r_i|\bar{\boldsymbol{r}}_{\neg i}) = \phi_i(r_i)\prod_{j\neq i}\psi_{ij}(r_i, \bar{r}_j) \times$$

$$\times \left[\prod_k \left(1 + \phi_k(h_{k,1})\frac{\psi_{ik}(h_{k,1}, r_i)}{\psi_{ik}(h_{k,1}, \bar{r}_i)}\prod_j \psi_{jk}(h_{k,1}, \bar{r}_j)\right)\right]$$

Thus we can compute all the $Z(r_i|\bar{\boldsymbol{r}}_{\neg i})$ in $\mathcal{O}(|\mathcal{S}|dN_u{}^2)$ time for all items rated by the user $u$.

Now we come to the gradient of the pseudo-likelihood

$$\partial\mathcal{L}^{PL}(\bar{\boldsymbol{r}}) = \sum_i \left(\partial \log Z(\bar{r}_i|\bar{\boldsymbol{r}}_{\neg i}) - \partial \log Z(\bar{\boldsymbol{r}}_{\neg i})\right) \quad (3)$$

Recall that $Z(\bar{\boldsymbol{r}}_{\neg i}) = \sum_{r_i} Z(r_i|\bar{\boldsymbol{r}}_{\neg i})$, we have

$$\begin{aligned}\partial \log Z(\bar{\boldsymbol{r}}_{\neg i}) &= \frac{1}{Z(\bar{\boldsymbol{r}}_{\neg i})}\sum_{r_i} \partial Z(r_i|\bar{\boldsymbol{r}}_{\neg i})\\ &= \frac{1}{Z(\bar{\boldsymbol{r}}_{\neg i})}\sum_{r_i} Z(r_i|\bar{\boldsymbol{r}}_{\neg i})\partial \log Z(r_i|\bar{\boldsymbol{r}}_{\neg i})\\ &= \sum_{r_i} P(r_i|\bar{\boldsymbol{r}}_{\neg i})\partial \log Z(r_i|\bar{\boldsymbol{r}}_{\neg i})\end{aligned}$$

Thus Eq.3 reduces to $\partial\mathcal{L}^{PL}(\bar{\boldsymbol{r}}) =$

$$\begin{aligned}&\sum_i \left(\sum_{r_i}\{\mathbb{I}[r_i = \bar{r}_i] - P(r_i|\bar{\boldsymbol{r}}_{\neg i})\}\partial \log Z(r_i|\bar{\boldsymbol{r}}_{\neg i})\right)\\ &= \sum_i \left(\sum_{r_i} D(r_i|\bar{\boldsymbol{r}}_{\neg i})\partial \log Z(r_i|\bar{\boldsymbol{r}}_{\neg i})\right)\end{aligned}$$

where $\mathbb{I}[.]$ is an identity function and $D(r_i|\bar{\boldsymbol{r}}_{\neg i}) = \mathbb{I}[r_i = \bar{r}_i] - P(r_i|\bar{\boldsymbol{r}}_{\neg i})$.

Let us consider $\partial \log Z(r_i|\bar{\boldsymbol{r}}_{\neg i})$. It is known that

$$\begin{aligned}\frac{\partial \log Z(r_i|\bar{\boldsymbol{r}}_{\neg i})}{\partial \alpha_k} &= P(h_{k,1}|r_i, \bar{\boldsymbol{r}}_{\neg i})\\ \frac{\partial \log Z(r_i|\bar{\boldsymbol{r}}_{\neg i})}{\partial \beta_{ja}} &= f_a(r_i)\mathbb{I}[i = j]\\ \frac{\partial \log Z(r_i|\bar{\boldsymbol{r}}_{\neg i})}{\partial \gamma_{jka}} &= \begin{cases} P(h_{k,1}|r_i, \bar{\boldsymbol{r}}_{\neg i})f_a(r_i) & \text{for } i = j\\ P(h_{k,1}|r_i, \bar{\boldsymbol{r}}_{\neg i})f_a(\bar{r}_j) & \text{for } i \neq j\end{cases}\end{aligned}$$

where $P(h_{k,1}|r_i, \bar{\boldsymbol{r}}_{\neg i})$ is

$$\frac{\phi_k(h_{k,1})\psi_{ik}(h_{k,1}, r_i)\prod_j \psi_{jk}(h_{k,1}, \bar{r}_j)}{\psi_{ik}(h_{k,1}, \bar{r}_i) + \phi_k(h_{k,1})\psi_{ik}(h_{k,1}, r_i)\prod_j \psi_{jk}(h_{k,1}, \bar{r}_j)}$$

Finally, we need to sum over all the visible nodes as in Eq.3

$$\begin{aligned}\frac{\partial\mathcal{L}^{PL}(\bar{\boldsymbol{r}})}{\partial \alpha_k} &= \sum_i\sum_{r_i} D(r_i|\bar{\boldsymbol{r}}_{\neg i})P(h_{k,1}|r_i, \bar{\boldsymbol{r}}_{\neg i})\\ \frac{\partial\mathcal{L}^{PL}(\bar{\boldsymbol{r}})}{\partial \beta_{ja}} &= \sum_{r_j} D(r_j|\bar{\boldsymbol{r}}_{\neg j})f_a(r_j)\\ \frac{\partial\mathcal{L}^{PL}(\bar{\boldsymbol{r}})}{\partial \gamma_{jka}} &= \begin{cases}\sum_{r_j} D(r_j|\bar{\boldsymbol{r}}_{\neg j})P(h_k|r_j, \bar{\boldsymbol{r}}_{\neg j})\Delta f_a(r_j)\\ +f_a(\bar{r}_j)\frac{\partial\mathcal{L}(\bar{\boldsymbol{r}})}{\partial \alpha_k}\end{cases}\end{aligned}$$

where $\Delta f_a(r_j) = f_a(r_j) - f_a(\bar{r}_j)$.

## A.2 BMs WITH GAUSSIAN RATINGS

Since ratings are sometimes provided in a numerical scale, they can be approximated by continuous variables, as suggested in (Hofmann, 2004). The energy of the system is given as

$$E(\boldsymbol{h}, \boldsymbol{r}) = \begin{cases} -\sum_k \alpha_k h_k - \sum_{i,k} \gamma_{ik} r_i h_k\\ +\sum_i \frac{r_i^2}{2} - \sum_i \beta_i r_i - \sum_{i,j\neq i} \lambda_{ij} r_i r_j\end{cases}$$

Here we assume that $P(r_i|\boldsymbol{r}_{\neg i}, h) = \mathcal{N}(\mu_i; 1)$, where $\mu_i = \beta_i + \sum_k \gamma_{ik} h_k + \sum_{j\neq i} \lambda_{ij} r_j$. Again, Gibbs sampling can be used for evaluating log-likelihood gradients in learning. In predicting new ratings, we can apply the mean-field approximation strategy described in Section 2.3, and compute the mode of the normal distribution $P(r_j|\bar{\boldsymbol{r}}, \boldsymbol{h})$, which is simply $\mu_j$

$$\mu_j = \beta_j + \sum_k \gamma_{jk} P(h_{k,1}|\bar{\boldsymbol{r}}) + \sum_i \lambda_{ij}\bar{r}_i$$

**Mean-field approximation to PL learning:**

Recall that the PL learning requires the conditional distribution $P(r_i|\bar{\boldsymbol{r}}_{\neg i})$, which is not Gaussian, making evaluation difficult. To turn it into a Gaussian, we can apply the mean-field approximation

$$\begin{aligned} P(r_i|\bar{\boldsymbol{r}}_{\neg i}) &= \sum_{\boldsymbol{h}} P(r_i, \boldsymbol{h}|\bar{\boldsymbol{r}}_{\neg i}) \\ &\approx Q(r_i|\bar{\boldsymbol{r}}_{\neg i}) \sum_{\boldsymbol{h}} \prod_k Q(h_k|\bar{\boldsymbol{r}}_{\neg i}) \end{aligned}$$

Further approximation $Q(h_k|\bar{\boldsymbol{r}}_{\neg i}) \approx P(h_k|\bar{\boldsymbol{r}})$ gives $Q(r_i|\bar{\boldsymbol{r}}_{\neg i}) \propto$

$$\exp\left( -\frac{r_i^2}{2} + \beta_i r_i + \sum_k P(h_{k,1}|\bar{\boldsymbol{r}})\gamma_{ik} r_i + \sum_{j\neq i} \lambda_{ij} r_i \bar{r}_j \right)$$

Thus $Q(r_i|\bar{\boldsymbol{r}}_{\neg i})$ is a Gaussian with mean $\mu_i = \beta_i + \sum_k P(h_k|\bar{\boldsymbol{r}})\gamma_{ik} + \sum_{j\neq i} \lambda_{ij}\bar{r}_j$.

Since the mean of a Gaussian is also its mode, PL learning can be approximately carried out by minimising the reconstruction error

$$\mathcal{E} = \frac{1}{2}\sum_i (\bar{r}_i - \mu_i)^2$$

Let $\epsilon_i = \bar{r}_i - \mu_i$. The gradients are

$$\begin{aligned} \frac{\partial \mathcal{E}}{\partial \beta_i} &= -\epsilon_i \\ \frac{\partial \mathcal{E}}{\partial \alpha_k} &= -P(h_{k,1}|\bar{\boldsymbol{r}}) \sum_i \epsilon_i \gamma_{ik} \\ \frac{\partial \mathcal{E}}{\partial \gamma_{ik}} &= -P(h_{k,1}|\bar{\boldsymbol{r}}) \left( \epsilon_i + P(h_{k,1}|\bar{\boldsymbol{r}}) r_i \sum_j \epsilon_j \gamma_{jk} \right) \\ &= -P(h_{k,1}|\bar{\boldsymbol{r}}) \left( \epsilon_i - r_i \frac{\partial \mathcal{E}}{\partial h_{k,1}} \right) \\ \frac{\partial \mathcal{E}}{\partial \lambda_{ij}} &= -\epsilon_i \bar{r}_j - \epsilon_j \bar{r}_i \end{aligned}$$